\documentstyle[aps,prb]{revtex}
\begin{document}
\draft

\title{Phonon Dispersion Effects and the Thermal Conductivity 
Reduction in GaAs/AlAs Superlattices}
\author{W. E. Bies, R. J. Radtke,$^*$ and H. Ehrenreich}
\address{Physics Department and
Division of Engineering and Applied Sciences,\\
Harvard University, Cambridge, Massachusetts~~02138}
\date{\today}

\maketitle

\begin{abstract}
The experimentally observed order-of-magnitude
reduction in the thermal conductivity along the growth axis of 
(GaAs)$_n$/(AlAs)$_n$ (or $n$x$n$)
superlattices is investigated theoretically for (2x2), (3x3) and 
(6x6) structures using an accurate model of the lattice dynamics.
The modification of the phonon dispersion relation due to the superlattice
geometry leads to flattening of the phonon branches and hence to lower 
phonon 
velocities. This effect is shown to account for a factor-of-three reduction
in the thermal conductivity with respect to bulk GaAs 
along the growth direction; the remainder
is attributable to a reduction in the phonon lifetime. The dispersion-related
reduction is relatively insensitive to temperature (100 $<$ T $<$ 300K) 
and  $n$. The phonon lifetime reduction is largest for the 2x2 
structures and consistent with greater interface scattering.  The thermal 
conductivity reduction is shown to be appreciably more sensitive to 
GaAs/AlAs force constant differences than to those associated with 
molecular masses.
\end{abstract}


\section{Introduction}

Superlattice structures have been proposed to be materials with a high
thermoelectric figure of merit {\it ZT}, for both in-plane \cite{HD}
and cross-plane \cite{MW,MSB} current flow. In the latter case, the
improvement in thermoelectric performance is attributable to a reduced
lattice thermal conductivity \cite{MW,MSB,REG} rather than to a higher 
electronic conductivity. Experimentally, a factor-of-ten reduction in
the component of lattice thermal conductivity along the growth axis,
{$\kappa_{\ell,zz}$}, is observed in GaAs/AlAs \cite{CM,CMRCPK} and 
Bi$_2$Te$_3$/Sb$_2$Te$_3$ \cite{VC,VSCSO} superlattices (SLs). 
Theoretically, 
the thermal conductivity of Si/Ge SLs has been studied previously by 
Hyldgaard and Mahan \cite{HM}  and by Chen \cite{Chen}.
Within the context of a very simplified 
model of the phonon dynamics,the calculations of Ref. 9  
were able to reproduce the factor-of-ten 
reduction in the thermal conductivity along the growth direction. By 
contrast, Chen's \cite{Chen} extensive work focussed on the role of thermal 
boundary resistance.  The present 
paper investigates the reduction associated with SL induced changes of 
the phonon dispersion based on a 
realistic, computationally intensive 
treatment of the phonon spectra and dynamics.

The origin of the observed reduction in thermal conductivity may be explained
qualitatively as follows. First, we note that, in the
experimental work of Capinski and Maris \cite{CM} and
Capinski {\it et al}~ \cite{CMRCPK} on (GaAs){$_n$}/(AlAs){$_n$} SLs
for {$n$} up to 40, the phonon mean free path inferred from the thermal
conductivity, heat capacity and Debye velocity is greater than 370 {\AA}
at all termperatures for which measurements exist, always large compared
to the size of the SL unit cell, 5.66 {\AA}. Thus, the
phonon transport lies in a regime where the SL phonon dispersion
relation and lifetime, and not those of the bulk constituents, 
determine the thermal conductivity. According to
the expression for the thermal conductivity derived from the phonon Boltzmann
equation in the relaxation-time approximation (see Eq. (6)), the thermal
conductivity depends on (1) a quantity representing the contribution of
the SL phonon dispersion relation, and (2) the lifetime $\tau$, which
contains phonon-phonon, interface and defect scattering effects. We shall
focus only on item (1), whose effect can be computed within our 
realistic, albeit complicated, lattice-dynamical model. 
The effect of the SL geometry is to introduce anticrossings and 
new gaps in the
phonon dispersion relation, when the magnitude of the phonon wavevector along 
the growth direction equals an integer multiple of {$\pi/d$}, where {$d$} is 
the period of the SL along the growth direction.
The consequent flattening of the phonon branches near the Brillouin zone edge
leads to a lowering of phonon velocities in the growth direction, and hence 
a reduction in thermal conductivity.

We describe in Section II the lattice-dynamical model for the SL, 
which is a generalization of the 11-parameter rigid-ion model of 
Kunc \cite{kunc1,kunc2}. It incorporates short-range interactions to
next nearest neighbors and the long-range Coulomb interaction.
The construction of the SL dynamical matrix is outlined in
Section II and in the Appendix.

The formalism is applied to a (GaAs){$_3$}/(AlAs){$_3$}
SL in Section III. 
The phonon dispersion relation will be seen to display the
flattening expected for a SL.
Critical points, especially at the high-symmetry points 
{$\Gamma$}, X and Z,
produce sharp peaks in the density of states in the SL.
Miniband formation and
anticrossings in the SL phonon dispersion relation lead to a
three-fold reduction, relative to bulk, in the contribution of the phonon 
dispersion relation to the thermal conductivity along the growth direction.
The present results contrast with those of Hyldgaard and Mahan \cite{HM}, who
found that, in their simplified model, the 
order-of-magnitude reduction of {$\kappa_{zz}/\tau$} was
attributable to effects related to the SL dispersion relation 
alone. In our more realistic treatment of the SL dynamics,
a significant three-fold reduction in the lifetime 
is also needed to explain the experimental reduction in 
{$\kappa_{zz}$} by a factor of ten. Finally, the sensitivity
of the decrease in {$\kappa_{zz}$} to differences in masses or force 
constants between the GaAs and AlAs layers is investigated; differences
in the force constants are found to play a markedly greater role in the
reduction of {$\kappa_{zz}$} than differences in mass.

Section IV is devoted to discussion and conclusions which are broadened by
examining the dependence of our results on the period {$n$} of
the (GaAs){$_n$}/(AlAs){$_n$} SL  for {$n=2,3$}
and 6. These results permit some discussion of interface effects on the 
phonon lifetime and a more detailed comparison of the present work 
with that of Refs. 9 and 10. 
The reduction in the contribution of the phonon dispersion
relation to the thermal conductivity of the SL relative to bulk
is computed, and found to be approximately independent of {$n$} for the 
small values considered here. 
The lifetime for both the SL and bulk was determined
using the experimental thermal conductivities of the SL
and of bulk GaAs of Capinski {\it et al}~ \cite{CMRCPK}.
The lifetimes are found to be smaller for the 2x2 SL and larger
and roughly equal for the 3x3 and 6x6 SLs, consistent with the 
presence of greater interface scattering in the 2x2 SLs.

\section{Formalism}

The lattice dynamics can be treated realistically via the 11-parameter
rigid-ion model of Kunc \cite{kunc1,kunc2}, which has been
successfully applied to zinc-blende-structure compounds. 
Consider (GaAs){${}_3$}/(AlAs){${}_3$}. In the SL unit cell,
consisting of a layer of GaAs above a layer of AlAs in the growth direction,
there will be 3 pairs of GaAs followed by 3 pairs of AlAs, or 3 Ga, 3 Al
and 6 As atoms in all, which may be indexed by {$\kappa=1,\ldots,12$}.
Letting
{$u_\alpha({\ell} \kappa)$} denote the displacement in the direction 
{$\alpha=x,y,z$} of the {$\kappa$}-th
atom in the {$\ell$}-th unit cell, plane-wave solutions of the form
\begin{equation}
u_\alpha({\ell} \kappa)=M_\kappa^{-1/2}
e^{i({\bf k} \cdot {\bf x}({\ell} \kappa) - \omega_j({\bf k})t)}
w_\alpha(\kappa|{\bf k} j)
\end{equation}
are assumed, where {${\bf x}({\ell}\kappa)$} is the equilibrium position
of the {$\kappa$}-th atom in the {$\ell$}-th unit cell, and
{$w_\alpha(\kappa|{\bf k}j)$} satisfies the secular equation
\begin{equation}
\omega_j({\bf k})^2 w_\alpha(\kappa|{\bf k}j) =
\sum_{\beta \kappa'} C_{\alpha \beta}(\kappa \kappa'|{\bf k}) 
w_\beta(\kappa'|{\bf k}j).
\end{equation}
The dynamical matrix {$\underline{C}$} reflects the interatomic 
force constants of the crystal. In the present rigid-ion model, the 
interatomic  forces are divided into (1) short-range forces extending to 
second nearest neighbors, and (2) the long-range Coulomb interaction. 
Accordingly, the dynamical matrix may be written
\begin{equation}
\underline{C}=\underline{C}_{sr}+\underline{C}_{\rm Coul}.
\end{equation}
As a result of symmetry of the zinc-blende structure, the short-range forces
to second nearest neighbors may be described by ten parameters for each 
material \cite{kunc2}.
For nearest and next nearest neighbor
interactions in the SL unit cell,
we employ the force constants determined for the constituent bulk materials 
separately, rotated by the appropriate point-group operation. Bulk GaAs and 
AlAs parameters are taken from the literature \cite{PPJS,RCC}. 
In the SL, bulk
parameters are used within each layer.  For the interface atoms and
Ga-Al bonds crossing the interface, 
we employ the average of the bulk parameters
following Ren {\it et al} \cite{RCC}. 
For the Coulomb interaction, the atoms are treated as point charges.
The Madelung sum, and its derivatives, are computed using the usual 
Ewald transformation \cite{BH}, which has been generalized here for 
SLs.
This is accomplished by separating the sum over the
spatial index {$\ell$} into a sum over layers normal to the
growth direction, {$\ell_\parallel$}, and a sum along the growth axis, 
{$\ell_z$}. A two-dimensional Ewald transformation is performed in each 
layer; these results are then summed over {$\ell_z$}.
The resulting expressions for the function {$\phi({\bf k},{\bf r})$} and its 
derivatives (see the Appendix) 
are similar to those that arise in the usual three-dimensional Ewald 
procedure; however, the definite integrals differ and must be 
performed numerically.
Detailed expressions for the Coulomb term are given in the Appendix.

The phonon Boltzmann equation in the relaxation-time approximation leads to
the following expression for the lattice thermal conductivity:
\begin{equation}
\kappa_{ij}=
\int {{d^3q} \over {(2\pi)^3}} \sum_\alpha
\hbar \omega^{(\alpha)}_{\bf q} 
{{\partial \omega^{(\alpha)}_{\bf q}} \over {\partial q_i}}
{{\partial \omega^{(\alpha)}_{\bf q}} \over {\partial q_j}}
{{dn(\omega^{(\alpha)}_{\bf q})} \over dT} 
\tau_{\rm ph}(\omega^{(\alpha)}_{\bf q},T)
\end{equation}
where $n(\omega _q^{(\alpha)})$ is the distribution function of the 
phonons, the sum is over branches {$\alpha$}, and 
{$\tau_{\rm ph}(\omega_{\bf q}^{(\alpha)},T)$} is the lifetime.
Eq. (4) can be written in terms of
\begin{equation}
\Sigma_{ij}(\omega)=
\int {{d^3q} \over {(2\pi)^3}} \sum_\alpha
\hbar \omega^{(\alpha)}_{\bf q} 
{{\partial \omega^{(\alpha)}_{\bf q}} \over {\partial q_i}}
{{\partial \omega^{(\alpha)}_{\bf q}} \over {\partial q_j}}
\delta(\omega-\omega^{(\alpha)}_{\bf q})
\end{equation}
as
\begin{equation}
\kappa_{ij}=\int d\omega (dn(\omega)/dT) \Sigma_{ij}(\omega)
\tau_{\rm ph}(\omega,T).
\end{equation}
This paper will focus on the SL effects on
the dispersion relation contained in {$\Sigma_{ij}(\omega)$}.
Relaxation-time effects associated for example with scattering from
interfaces, defects, umklapp processes, etc. are not considered explicitly. 
However, the results will be used to infer some of 
their properties.

\section{(G\lowercase{a}A\lowercase{s})$_3$/(A\lowercase{l}A\lowercase{s})$_3$ 
Superlattices}

We focus first on the (GaAs){${}_3$}/(AlAs){${}_3$} SL studied 
experimentally by Capinski and Maris \cite{CM}. Using a picosecond
pump-and-probe technique, they observed an order-of-magnitude reduction
in the thermal conductivity along the growth direction, {$\kappa_{zz}$},
relative to bulk GaAs.
The dispersion relation along the {$\Gamma$}X and {$\Gamma$}Z directions,
which was generated numerically according to the method described in
Section II, is shown in Fig. 1. The significance of the labelled features
will be explained in the discussion of Fig. 2. 
Because the SL unit cell contains three unit
cells each of bulk GaAs and bulk AlAs, respectively, arranged along the
growth axis, the edge of the SL Brillouin zone in the growth
direction, Z, is one-sixth as far from the center as it is in the in-plane
directions, X and Y.  As a result each of the six branches in the bulk
material is folded back six times along {$\Gamma$}Z, which is most easily
seen for the longitudinal acoustic mode in Fig. 1.
Bulk GaAs and bulk AlAs optical modes have no frequencies in common, 
and are therefore localized. This leads to (1) flat SL dispersion 
in optical modes,
and (2) localization of AlAs optical modes to the AlAs layer. The GaAs optical
modes are not localized, since they overlap with the acoustic modes as shown 
in Fig. 1 along {$\Gamma$}X.  Due to the non-analyticity of 
$\underline{C}_{\rm Coul}$ as $q\rightarrow 0$ in the SL, 
$\omega (\vert q \vert \rightarrow 0)$ differ along $\Gamma$X and 
$\Gamma$Z (See Ref. 14). 

The density of states {$\rho(\omega)$} versus frequency, computed using the
tetrahedral integration method as presented in MacDonald {\it et al}~
\cite{MVC}, is given in Fig. 2. Note the transverse-acoustic features
around {$\omega=80$} to 100 cm{${}^{-1}$}, the longitudinal-acoustic features
around 150 to 200 cm{${}^{-1}$}, the GaAs optical feature at
220 to 260 cm{${}^{-1}$},
and, separated in frequency at higher frequencies, the AlAs optical
feature at 330 to 400 cm{${}^{-1}$}.
Band gaps at the zone edge and anticrossings in Fig. 1 yield
critical points which, depending on the amount of {$\bf q$}-space at those
frequencies, produce sharp structure in the density of states.
As shown in Fig. 2, this structure in the density of states can 
be correlated with features in the dispersion relation, usually at the
{$\Gamma$} (for folded-back bands), X and Z points. The structures 
at {$\Gamma$}, X and Z labelled in Fig. 2 are identified with the 
corresponding features in Fig. 1. The strength of each feature depends 
on the integral of
{$\delta(\omega-\omega^{(\alpha)}_{\bf q})$} (cf. Fig. 2)
at that frequency, which will
be large if {$|{\bf v}|$} is small. Surprisingly, most of the peaks in the
density of states appear to be 
associated with the critical points at {$\Gamma$},
X and Z. No such fine structure in the density of states exists in 
bulk GaAs or bulk AlAs, as may be seen for instance in Patel {\it et al}~
\cite{PPJS} for GaAs. (The density of states for AlAs is not available in
the literature, but our calculations confirmed the absence of fine structure
for AlAs as well.)

Fig. 3 shows the results for {$\Sigma(\omega)$}, as defined in
Eq. (5), for bulk GaAs and the SL. Note that optical modes do
not contribute appreciably to {$\Sigma_{zz}(\omega)$} in the SL, an effect
of localization in the AlAs layers: flat dispersion leads to a vanishing
of {$\partial \omega^{(\alpha)}_{\bf q}/\partial q_z$}. The fine structure 
in the density of states is also correlated with that in 
{$\Sigma(\omega)$}: peaks in the density of states DOS {$\propto
\int dS/|{\bf v}|$} become dips in {$\Sigma \propto \int v^2dS/|{\bf v}|$}.
(Here, {$S$} denotes the surface of constant frequency {$\omega$} in the
Brillouin zone; it consists not only of the surface containing the critical
point, but also possibly of other surfaces at 
the same $\omega$ elsewhere in the zone.)
Acoustic modes in the SL contribute less than they would in bulk
because of the band-gap and anticrossing-induced reduction in {$v^2$}.
This leads to a three-fold reduction in {$\kappa_\ell/\tau$} at 300 K, 
determined here either by integrating Eq. (4) directly on a {$60\times 60\times 20$}
grid covering an irreducible wedge of the Brillouin zone, or by integrating
Eq. (6). Experimentally, Capinski and Maris \cite{CM} found a 
ten-fold reduction factor
for (GaAs){${}_3$}/(AlAs){${}_3$}. The full reduction factor
is a product of the reduction due to {$\Sigma$} and that
due to {$\tau$}. Assuming {$\tau$} to be constant 
at any given temperature, it 
can be found by requiring equality in Eq. (4) or (6) to the experimental 
value for the thermal conductivity in bulk or SL, respectively, 
if the appropriate 
dispersion relation  
is used in computing {$\kappa_{zz}/\tau$}. The  
lifetime in bulk versus lifetime in SL, as well as the layer-width
dependence of the results, will be discussed below. Finally, we note that 
{$\Sigma_{zz}<\Sigma_{xx}$} in the SL for the simple physical
reason that band flattening along the {$q_z$}-direction affects
{$\partial \omega/\partial q_z$} more than {$\partial \omega/\partial q_x$}.

The reduction in thermal conductivity due to {$\Sigma$} can be
understood by means of an easily visualized picture in {\bf q}-space.
In bulk, the quantity {$q_x \sum_\alpha (dn(\omega^{(\alpha)}_{\bf q})/dT) 
\omega^{(\alpha)}_{\bf q} (v^{(\alpha)}_{{\bf q},z})^2$} 
represents the contribution
of the phonon dispersion relation in Eq. (4). In a range {$\Delta q_x$}
of the integrand it is weighted by {$q_x$} 
because, the dispersion relation being rotationally symmetric in the
{$(q_x,q_y)$} plane, we may integrate around the circle of radius {$q_x$}
to yield a properly weighted function of {$q_x$} and {$q_z$} alone.
The quantity is plotted in Fig. 4(a),(b),(c) (bold line) together with
the corresponding values for GaAs (light solid line) as a function of
$q_z$ for three values of $q_x$. The SL
contribution is reasonably localized in {$q_z$}. To gain physical 
insight, we replace it by the
dashed rectangles (of equal area). 
This approximate localization is related to the reduction at
{$q_z \approx 0$} and {$\pi/d$}
due to miniband formation (that is, flattening of {$\omega$} vs. {$q$})
and has dips from anticrossings, as in Fig. 4(a), for instance. 
The dependence on {$q_x$} can then be summarized by the
density plot in Fig. 4(d), comparing the SL on the left to
bulk GaAs on the right. For GaAs the equivalent rectangles extend over the 
entire range of {$q_z$} because there is no localization due to band 
flattening at the zone edge as in the case of the SL. 
The shading indicates the weight of each increment {$\Delta q_x$}.
We find that points around {$q_x=\pi/d$} and
at the zone edge ({$q_x=6\pi/d$}) contribute most to heat transport. The
latter is due to the effect of the weighting by {$q_x$} in the annular 
integration. In addition, this weighting causes the contribution from points 
near the origin to be very small.
A simple numerical estimate leads to a reduction in {$\Sigma$} to
34\%, which is to be compared to 36{\%} in the exact calculation. 

Finally, we discuss the sensitivity of variation of the mass differences,
force constants and effective ionic charge 
{$e^*$} between layers on the thermal conductivity. 
This manifests itself through variations of the zone edge gap and hence
the group velocity. This effect has been studied
by interpolating {$e^*$}, the force constants $K$, and cation mass
$M$ between their AlAs and their GaAs values
in the GaAs layer. Here {$K$} refers collectively to the 10 Kunc parameters 
describing interatomic forces.  Explicitly, starting
from (AlAs){${}_6$} we (i) vary {$e^{*2}$} and {$M$} 
linearly in three neighboring AlAs positions to make a (AlAs)$_3$/(GaAs)$_3$ SL.
In the case of the masses we put 
{$M=(1-\alpha)M_{\rm Al}+\alpha M_{\rm Ga}$} so that 
at {$\alpha=0$} and 1 correspond respectively to  Al and  Ga;
(ii) vary {$e^{*2}$} and force constants {$K$} linearly in {$\alpha$} 
in the same way;  and 
(iii) vary {$e^{*2}$}, and both {$M$} and {$K$} linearly in 
{$\alpha$}.
The results are given in Fig. 5. The thermal conductivity
is seen to be far more sensitive to the variation of
force constants than the variation in mass.
The variation of the
force constants alone produces band flattening which reduces
the thermal conductivity by about 60\%. The additional
flattening due to changing masses leads to only a few percent additional
reduction.

\section{Discussion and Conclusions}

Before turning to the final results, we emphasize again that this paper is 
primarily concerned with the effects of superlattice induced changes of the 
phonon dispersion on the lattice thermal conductivity.  Since a SL is a 
perfect crystal, the ideal structures under discussion here automatically 
include the coherent effects associated with perfect interfaces.  The 
importance of including such effects was first pointed out by Hyldgaard and 
Mahan \cite{HM} 
in connection with a simple, highly idealized model for Si/Ge SLs.  
Extensive work by Chen \cite{Chen} focussed on 
diffuse interface scattering of phonons.  His  model assumes  SL layers 
sufficiently thick that the phonon spectrum in each layer corresponds to 
that of the bulk.  A mixture of spectral and diffuse interface processes is 
found to be sufficient to explain the observed experimental reduction of the 
SL thermal conductivity relative to bulk.  As already pointed out, the more 
realistically modeled results of the present paper for $\kappa _\ell /\tau$ 
lead to a three fold reduction without any assumptions about the SL phonon 
scattering mechanisms.  Our results must therefore be viewed as complementary 
to those of Chen \cite{Chen}.  Taken together, 
they suggest that a combination of phonon spectral 
changes and imperfect interfaces can account adequately for the observed 
reduction.

The present calculations permit some statements concerning lifetime effects 
from  the dependence of the SL results on layer width.  This dependence was 
studied numerically for
2x2, 3x3 and 6x6 GaAs/AlAs SLs. The results
for 
\begin{equation}
\bar{\Sigma}_{zz} \equiv \kappa_{zz}/\tau =
\int {{d^3q} \over {(2\pi)^3}} \sum_\alpha
\hbar \omega^{(\alpha)}_{\bf q} 
{{\partial \omega^{(\alpha)}_{\bf q}} \over {\partial q_i}}
{{\partial \omega^{(\alpha)}_{\bf q}} \over {\partial q_j}}
{{dn(\omega^{(\alpha)}_{\bf q})} \over dT} 
\end{equation}
are given in Table I. {$\tau$} is assumed constant. 
Given an experimental value for {$\kappa_{zz}$} and the calculated value
of {$\bar{\Sigma}_{zz}$}, the lifetimes listed in Table I are found as
{$\tau=\kappa_{zz}/\bar{\Sigma}_{zz}$}; only the lifetime itself is given 
in Table I and not {$\bar{\Sigma}_{zz}$}, but the ratios 
$\bar{\Sigma}_{zz}$(SL)/$\bar{\Sigma}_{zz}$(bulk) are listed because
we are interested in 
\begin{equation}
{\tau_{\rm SL} \over \tau_{\rm bulk}} = {{\bar{\Sigma}_{zz}({\rm bulk})}
\over {\bar{\Sigma}_{zz}({\rm SL})}} {{\kappa_{zz}^{\rm SL}({\rm expt})} \over
{\kappa_{zz}^{\rm bulk}({\rm expt})}}.
\end{equation}
The SL {$\bar{\Sigma}_{zz}$}
is found to have a value about 40{\%} of the bulk, and to be relatively 
insensitive to the SL period and temperature.
For larger SL periods, there is more folding back, but this
is compensated by the smaller size of the gaps at the zone edge, which
is found in the computed {$\Gamma$}Z dispersion relations to scale
inversely with the SL period. The two effects balance,
resulting in an approximately constant reduction factor.
The phonon lifetimes for the 2x2, 3x3 and 6x6 SLs are also
given in Table I. The bulk lifetime is the same to within 2{\%} for the
different SL periods, as it must be. The ratio of SL
to bulk lifetimes is significantly smaller for the 2x2 SLs at each
temperature while it is larger and roughly the same for the 3x3 and 6x6
SLs. Thus, the reduction in thermal conductivity may be
divided into a dispersive part which is insensitive to {$n \times n$}
and a scattering part which is sensitive to interface scattering. The 
results for the 2x2 SL are possibly associated with the experimental 
difficulties associated with achieving sufficiently perfect interfaces 
for small {$n$}.

The present calculations imply:  (1) that a similar reduction in the contribution 
of the SL phonon disperion relation to transport in the growth direction, 
and perhaps in $\kappa _{zz}$ itself, may be expected in any SL with 
similar mass or force-constant differences between layers.  (2) If the 
lifetime is reduced in SLs by increased umklapp scattering, as suggested by 
Ren and Dow$^{13}$, then the present calculations give an upper bound on 
the SL $\kappa _{zz}$ and a lower bound on the increase in $ZT$ in a SL 
relative to bulk.

\acknowledgements

We wish to thank Dr. E. Runge for stimulating discussions. 
This work was supported by DARPA through ONR Contract No.~N00014-96-1-0887
and NSF Grant Che9610501.

\appendix
\section{}

This Appendix presents the detailed formulae for the Coulomb
part of the dynamical matrix, derived using the Ewald procedure as
described in the text of the paper. Letting {$\kappa$} label the
atoms of the SL unit cell, {$M_\kappa$} be the mass of
the {$\kappa$}-th atom and {${\bf x}(\kappa\kappa')$}
be the separation vector from the {$\kappa$}-th atom to the {$\kappa'$}-th 
atom, we have, for {$\kappa \ne \kappa'$},
\begin{equation}
C^{\rm Coul}_{\alpha\beta}(\kappa\kappa'|{\bf k}) =
-{{e_\kappa e_{\kappa'}} \over \sqrt{M_{\kappa}M_{\kappa'}}}
e^{i{\bf k}\cdot{\bf x}(\kappa'\kappa)}
{{\partial^2 \phi({\bf k},{\bf r})} \over 
{\partial r_\alpha \partial r_\beta}}|_{{\bf r}={\bf x}(\kappa'\kappa)}
\end{equation}
where {$\phi({\bf k},{\bf r})$} is by definition
\begin{equation}
\phi({\bf k},{\bf r})=\sum_\ell {{e^{i{\bf k}\cdot{\bf x}(\ell)}} \over
{|{\bf x}(\ell)+{\bf r}|}},
\end{equation}
{${\bf x}(\ell)$} being the position of the {$\ell$}-th unit cell.
The result of the Ewald procedure adapted to the superlattice is that 
{$\phi({\bf k},{\bf r})$} can be written in the form
\begin{eqnarray}
\phi({\bf k},{\bf r}) &=& R \sum_\ell H(|{\bf x}(\ell)+{\bf r}|R) 
e^{i{\bf k}\cdot{\bf x}(\ell)} \nonumber \\
&+& {{2\sqrt{\pi}} \over v_\parallel} \sum_{h_\parallel,\ell_z}
{2 \over {|\tau(h_\parallel)+{\bf k}_\parallel|}}
I(\infty,{{|\tau(h_\parallel)+{\bf k}_\parallel|} \over {2R}},
{{|\tau(h_\parallel)+{\bf k}_\parallel||{\bf x}(\ell_z)+z|} \over 2}) 
e^{-i(\tau(h_\parallel)+{\bf k}_\parallel)\cdot {\bf r}_\parallel+
i k_z \hat{z} \cdot {\bf x}(\ell_z)}
\end{eqnarray}
where {$R$} is an arbitrary cutoff (we always take {$R=3/a_0$}),
\begin{equation}
H(x)={{2/\sqrt{\pi}} \over x} \int_x^\infty e^{-x'^2} dx',
\end{equation}
{$v_\parallel$} is the area of the unit cell in the superlattice plane,
{$\ell_z$} labels the layers perpendicular to the growth ({$z$}) axis,
{$h_\parallel=(h_x,h_y)$} labels the cells located at reciprocal lattice
vectors {$\tau(h_\parallel)$} in each plane, 
{${\bf k}=({\bf k}_\parallel,k_z)$}, and
\begin{equation}
I(\alpha,\beta,\gamma)=\int_\beta^\alpha e^{-v^2-\gamma^2/v^2} dv.
\end{equation}

For {$\kappa=\kappa'$} we have
\begin{equation}
C^{\rm Coul}_{\alpha\beta}(\kappa\kappa|{\bf k})
= {{e_\kappa} \over {M_\kappa}}
\left[ -e_\kappa \sum_{\ell \ne 0} e^{i{\bf k}\cdot{\bf x}(\ell)}
\left( {{\partial^2 r^{-1}} \over {\partial r_\alpha \partial r_\beta}}
\right)_{{\bf r}={\bf x}(\ell)} \\
+ \sum_{\ell'\kappa' \ne 0\kappa} e_{\kappa'} \left( 
{{\partial^2 r^{-1}} \over {\partial r_\alpha \partial r_\beta}}
\right)_{{\bf r}={\bf x}(\ell'\kappa',0\kappa)} \right]. 
\end{equation}
The first term on the right is similar to the above expressions in
Eqs. (A1)-(A3) but for the {$\ell=0$} in Eq. (A3) term one substitutes
{$(4/3\sqrt{\pi})\delta_{\alpha\beta}$}. The second term is given by
\begin{equation}
{{\partial^2} \over {\partial r_\alpha \partial r_\beta}}
\left[ I_0 + I_1 + I_2 \right]_{{\bf r}=0}
\end{equation}
with
\begin{equation}
I_0=
R e_\kappa H^0(|{\bf r}|R),
\end{equation}
\begin{equation}
I_1= R \sum_{\ell'\kappa'\ne 0\kappa} e_{\kappa'}
H(|{\bf x}(\ell')+{\bf x}(\kappa'\kappa)+{\bf r}|R)
\end{equation}
and 
\begin{equation}
I_2 = {{2\sqrt{\pi}} \over {v_\parallel}} 
\sum_{h_\parallel, \ell_z, \kappa'} e_{\kappa'} 
{2 \over {|\tau(h_\parallel)|}}
I(\infty,{{|\tau(h_\parallel)|} \over {2R}},{{|\tau(h_\parallel)|
|{\bf x}(\ell_z)+\hat{z}\cdot{\bf x}(\kappa'\kappa)|} \over 2})
e^{-i\tau(h_\parallel)\cdot({\bf x}(\kappa'\kappa)+{\bf r})}
\end{equation}
where
\begin{equation}
H^0(x)=-{{2/\sqrt{\pi}} \over x} \int_0^x e^{-x'^2} dx'.
\end{equation}
A similar, but not identical, approach was used in Ref.~15.


\newpage

\begin{table}

\caption{Reduction in {$\bar{\Sigma}_{zz}=\kappa_{zz}/\tau$} 
relative to bulk GaAs
for 2x2, 3x3 and 6x6 SLs at {$T=100, 200$} and 300 K,
and SL phonon lifetimes deduced from Eq. (8) for the
2x2, 3x3 and 6x6 SLs at each temperature.}

\begin{tabular}{cccccc}
   &   experimental            & theoretical & &  \\
SL & 
$\kappa_{zz}^{\rm SL}$ (W/cm K)$^a$ &
${{\bar{\Sigma}_{zz}({\rm SL})} \over {\bar{\Sigma}_{zz}({\rm bulk})}}$ &
{$\tau_{\rm bulk}$} (ps) & 
$\tau_{\rm SL}$ (ps) &
$\tau_{\rm SL}/\tau_{\rm bulk}$  \\
\hline
T=300 K, $\kappa_{zz}^{\rm bulk}$=0.45 W/cm K $^b$\\
2x2 & 0.040 & 0.38 & 36.8 & 8.7  & 0.24 \\
3x3 & 0.068 & 0.36 & 37.2  & 15 & 0.42 \\
6x6 & 0.053 & 0.34 & 37.2 & 13  & 0.35 \\
\hline
T=200 K, $\kappa_{zz}^{\rm bulk}$=0.64 W/cm K $^b$\\
2x2 & 0.055 & 0.38 & 55.2 & 12 & 0.22 \\
3x3 & 0.090 & 0.39 & 56.1 & 20 & 0.36 \\
6x6 & 0.072 & 0.35 & 55.7 & 18 & 0.32 \\
\hline
T=100 K, $\kappa_{zz}^{\rm bulk}$=2.0 W/cm K $^b$\\
2x2 & 0.065 & 0.40 & 222 & 18 & 0.08 \\
3x3 & 0.110 & 0.42 & 227 & 30 & 0.14 \\
6x6 & 0.096 & 0.37 & 222 & 29 & 0.13 \\
\end{tabular}
$^a$ Experimental values for GaAs/AlAs reported by 
Ref.~{\protect\onlinecite{CMRCPK}}.

$^b$ Experimental value for GaAs listed in Ref.~{\protect\onlinecite{CM}}.

\end{table}

\newpage
\begin{center}
FIGURES
\end{center}
\vskip 0.3in

FIG. 1. (GaAs)$_3$/(AlAs){$_3$} SL dispersion relation along the
{$\Gamma$}X={$(2\pi/a_0,0,0)$} and {$\Gamma$}Z={$(0,0,\pi/3a_0)$} directions; 
$a_0$ is the conventional GaAs unit cell size.
The labels $a-w$ are defined in Fig. 2.

FIG. 2. Phonon density of states for the (GaAs)$_3$/(AlAs){$_3$} SL,
with labelled critical points identified with features in the dispersion
relation in Fig. 1.

FIG. 3. The transport quantities 
{$\Sigma_{xx}(\omega)$} and {$\Sigma_{zz}(\omega)$}, defined by
Eq. (5) in the text, for bulk GaAs (solid line) and the 
(GaAs)$_3$/(AlAs){$_3$} SL (dashed and bold lines).

FIG. 4. The transport quantity 
{$q_x(dn/dT)\sum_\alpha \omega_{\bf q}^{(\alpha)} 
(v_{{\bf q},z}^{(\alpha)})^2$}
for {$0 \le q_z \le \pi/d$} at fixed {$q_x$}, for (a) {$q_x=2\pi/d$},
(b) {$q_x=4\pi/d$} and (c) {$q_x=6\pi/d$}; solid line:  bulk GaAs; 
bold line: (GaAs)$_3$/(AlAs){$_3$} SL.
(d) Density plot in the {$(q_x,q_z)$
plane whose shading indicates the weight of each increment $\Delta q_x$
along $q_x$ to the  value of this
transport quantity for the (GaAs)$_3$/(AlAs){$_3$} SL and bulk GaAs.

FIG. 5. Dependence of {$\kappa_{zz}$} as {$e^{*2}$} and the mass {$M$} 
(solid line), the spring constants {$K$} (long-dashed line) and both {$M$}
and {$K$} (short-dashed line) are interpolated between their AlAs ({$\alpha=0$})
and GaAs values ({$\alpha=1$}) for the atoms in three contiguous layers
in what is initially (AlAs){$_6$} at {$\alpha=0$}. 


\begin{references}

\item[*]{Present address:  Schlumberger, Sugar Land Product Center, 
110 Schlumberger Dr., MD110-4, Sugar Land, TX  77478.}

\bibitem{HD}
L. D. Hicks and M. S. Dresselhaus, {\it Phys. Rev. B} {\bf 47}, 12727 (1993).

\bibitem{MW}
G. D. Mahan and L. M. Woods, {\it Phys. Rev. Lett.} {\bf 80}, 4016 (1998).

\bibitem{MSB}
G. D. Mahan, J. O. Sofo and M. Bartkowiak, {\it J. Appl. Phys.} {\bf 83},
4683 (1998).

\bibitem{REG}
R. J. Radtke, H. Ehrenreich and C. H. Grein, {\it J. Appl. Phys} {\bf 86}, 
3195 (1999).

\bibitem{CM}
W. S. Capinski and H. J. Maris, {\it Physica B} {\bf 219}, 699 (1996).

\bibitem{CMRCPK}
W. S. Capinski, H. J. Maris, T. Ruf, M. Cardona, K. Ploog and D. S. Katzer,
{\it Phys. Rev. B} {\bf 59}, 8105 (1999).

\bibitem{VC}
R. Venkatasubramanian and T. Colpitts, Mat. Res. Soc. Symp. Proc. {\bf 478},
73 (1997).

\bibitem{VSCSO}
R. Venkatasubramanian, E. Sivola, T. Colpitts K. Stokes and B. O'Quinn, 
Mat. Res. Soc. Symp. Proc. {\bf 545} (to appear).

\bibitem{HM}
P. Hyldgaard and G. D. Mahan, {\it Phys. Rev. B} {\bf 56}, 10754 (1997).

\bibitem{Chen}
G. Chen, {\it J. Heat Transfer} {\bf 119}, 220 (1996); {\it Phys. Rev. B} 
{\bf 57}, 14958 (1998).

\bibitem{kunc1}
K. Kunc, {\it Ann. Phys.,} 1973-1974, t. 8, pp. 319-401.

\bibitem{kunc2}
K. Kunc, M. Balkanski, and M. A. Nusimovici, {\it Phys. Rev. B} {\bf 12}, 4346
(1975).

\bibitem{RD}
S. Y. Ren and J. D. Dow, {\it Phys. Rev. B} {\bf 25}, 3750 (1982).

\bibitem{PPJS}
C. Patel, T. J. Parker, H. Jamshidi and W. F. Sherman, {\it Phys. Stat. Sol.
(b)} {\bf 122}, 461 (1984).

\bibitem{RCC}
S.-F. Ren, H. Chu and Y.-C. Chang, {\it Phys. Rev. B.} {\bf 37}, 8899 (1988).

\bibitem{BH}
M. Born and K. Huang, {\it Dynamical Theory of Crystal Lattices} 
(Oxford, 1954), pp. 248-255.

\bibitem{MVC}
A. H. MacDonald, S. H. Vosko and P. T. Coleridge, {\it J. Phys. C}
{\bf 12}, 2991 (1979).

\end{references}
\end{document}